\documentclass[aps,prb,superscriptaddress,reprint,longbibliography,amsmath,amssymb]{revtex4-2}
\usepackage{graphicx}
\usepackage{dcolumn}
\usepackage{bm}
\usepackage{physics}
\usepackage{amsmath}
\usepackage{amssymb}
\usepackage{lipsum}
\usepackage{wrapfig}
\usepackage{booktabs}
\usepackage{array}
\usepackage{float}
\usepackage{multirow}
\usepackage{braket}
\usepackage[colorlinks,linkcolor=blue,urlcolor=blue,anchorcolor=blue,citecolor=blue]{hyperref}
\usepackage{placeins}
\usepackage{cancel}
\usepackage{xcolor}

\begin{document}
\newcommand{\ud}{$\mathrm{d}$}
	
        \title{Anderson Lattice in Incommensurate $\bf{Nb_3Cl_8}$/Graphene van der Waals Heterostructures}
	
	
    \author{Yuchen Gao}
    \affiliation{State Key Laboratory for Artificial Microstructure $\rm{\&}$ Mesoscopic Physics and Frontiers Science Center for Nano-Optoelectronics, School of Physics, Peking University, Beijing 100871, China}

    \author{Wenjie Zhou}
    \affiliation{State Key Laboratory for Artificial Microstructure $\rm{\&}$ Mesoscopic Physics and Frontiers Science Center for Nano-Optoelectronics, School of Physics, Peking University, Beijing 100871, China}
    
    \author{Fan Yang}
    \affiliation{State Key Laboratory for Artificial Microstructure $\rm{\&}$ Mesoscopic Physics and Frontiers Science Center for Nano-Optoelectronics, School of Physics, Peking University, Beijing 100871, China}

    \author{Zhijie Ma}
    \affiliation{University of Chinese Academy of Sciences, Beijing 100049, China}
    \affiliation{Songshan Lake Materials Laboratory, Dongguan, Guangdong 523808, China}

    \author{Hansheng Xu}
    \affiliation{State Key Laboratory for Artificial Microstructure $\rm{\&}$ Mesoscopic Physics and Frontiers Science Center for Nano-Optoelectronics, School of Physics, Peking University, Beijing 100871, China}
    
    \author{Xinyue Huang}
    \affiliation{State Key Laboratory for Artificial Microstructure $\rm{\&}$ Mesoscopic Physics and Frontiers Science Center for Nano-Optoelectronics, School of Physics, Peking University, Beijing 100871, China}
    \affiliation{Academy for Advanced Interdisciplinary Studies, Peking University, Beijing 100871, China}

    \author{Kenji Watanabe}
    \affiliation{Research Center for Electronic and Optical Materials National Institute for Materials Science 1-1 Namiki, Tsukuba 305-0044, Japan}
    
    \author{Takashi Taniguchi}
    
    \affiliation{Research Center for Materials Nanoarchitectonics National Institute for Materials Science 1-1 Namiki, Tsukuba 304-0044, Japan}
    
    \author{Youguo Shi}
    \email{ygshi@iphy.ac.cn}
    \affiliation{University of Chinese Academy of Sciences, Beijing 100049, China}
    \affiliation{Songshan Lake Materials Laboratory, Dongguan, Guangdong 523808, China}
     
    \author{Yu Ye}
    \email{ye\_yu@pku.edu.cn}
    \affiliation{State Key Laboratory for Artificial Microstructure $\rm{\&}$ Mesoscopic Physics and Frontiers Science Center for Nano-Optoelectronics, School of Physics, Peking University, Beijing 100871, China}
    \affiliation{Collaboration International Center of Quantum Matter, Beijing 100871, China}
    \affiliation{Liaoning Academy of Materials, Shenyang 110167, China}
     
	%
	
	
 
\begin{abstract}

 {The periodic Anderson model}, traditionally realized in rare-earth compounds with limited tunability, have hindered systematic exploration of correlated quantum phenomena. Here, we introduce a strategy for  {realizing and }engineering  {this model} in incommensurate van der Waals heterostructures by coupling a Mott insulator (Nb$_3$Cl$_8$) with itinerant electrons (from monolayer graphene), circumventing strict lattice-matching requirements. Through magnetotransport and slave spin mean-field calculations, we demonstrate the hybridization gap ($\Delta\approx30$ meV), gate-tunable metal-insulator transition, and band-selective electron effective mass enhancement, hallmarks of Kondo coherence. The heterostructure exhibits a nearly order-of-magnitude enhancement in the effective electron mass between hybridized and conventional graphene-like regimes, alongside in-plane magnetic field-induced metal-insulator transitions. Top gate-temperature phase mapping reveals competing correlated states, including insulating and hidden-order phases. This work establishes an electrically tunable van der Waals platform for studying correlated states generated by coupling a Mott-insulating layer to an itinerant-electron system, providing a materials route for exploring low-dimensional correlated quantum phases.

\end{abstract}

\maketitle

\textit{Introduction-}- { {Periodic Anderson model (PAM) is the minimum model that describes coupling between localized and itinerant electrons. It constitutes a paradigmatic platform for studying strongly correlated quantum matter, whose realization introduces quantum criticality, non-Fermi-liquid behavior, topological quasiparticles, and unconventional superconductivity  \cite{stewart2001non, dzero2016topological, stockert2011magnetically, jiao2020chiral}. {In materials described by the PAM}, the hybridization between localized moments and itinerant electrons produces strongly renormalized quasiparticles. Despite decades of progress,   {the experimental exploration of this model} are typically based on rare-earth or actinide materials with limited tunability and substantial structural complexity, making it difficult to systematically control the localized and itinerant degrees of freedom  \cite{maksimovic2022evidence, paschen2004hall}. Recent advances in van der Waals (vdW) heterostructures have introduced artificial platforms for  Anderson-lattice physics, including moir\'e-engineered MoTe$_2$/WSe$_2$  \cite{zhao2023gate} and commensurate 1T/1H-TaS$_2$ heterostructures  \cite{vavno2021artificial}.} VdW heterostructures enable \textit{in situ} control of Kondo lattice interactions and quantum criticality.  {However, these systems still rely on specific lattice matching or material combinations. An adaptable heterostructure platform that decouples the localized and itinerant components while retaining electrostatic tunability would therefore provide a useful route for exploring correlated interfacial states in a broader materials space.}

\begin{figure*}[ht!]
\centering 
\includegraphics{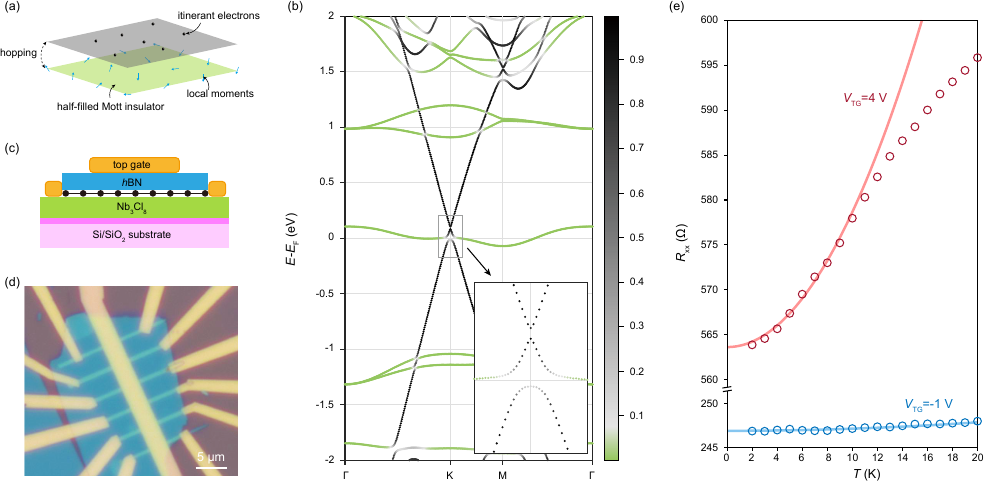}
\caption{\textbf{Structural and electronic properties of the Nb$_3$Cl$_8$/MLG heterostructure.}
 \textbf{a,} Schematic of the heterostructure. Monolayer graphene (MLG, gray) hosts itinerant carriers, while the Nb$_3$Cl$_8$ (green) layer forms a Mott insulator with localized moments at Nb$_3$ clusters. Interlayer electron hopping realizes the periodic Anderson model.
 \textbf{b,} DFT-calculated band structure of Nb$_3$Cl$_8$/MLG heterostructure. Inset: the zoomed DFT band structure near the Fermi level. The color bar denotes the carbon orbital weight.
 \textbf{c-d,} Side view (c) and optical image (d) of the dual-gated Hall bar device.
 \textbf{e,} Temperature-dependent resistance at $V_{\rm{TG}}=-1$ V and  $V_{\rm{TG}}=4$ V. Both curves follow $R=R_0+AT^2$ (solid lines), confirming Fermi liquid nature. The contrasting residual resistance $R_0$ and coefficient $A$ highlight distinct electronic states.}
\label{F1}
\end{figure*}

We explicitly demonstrate the  {realization of the PAM} within a distinctive configuration in which a layer of itinerant electrons is stacked upon another flat-band Mott insulator layer. In this unique set-up, the nearly half-filled flat band accommodates localized electrons, while the interlayer hopping facilitates the hybridization of itinerant and localized electron orbits, as elucidated in Fig. \ref{F1}a. For our experimental framework, Nb$_3$Cl$_8$ serves as the Mott insulator, distinguished by a prominent half-filled flat band with a bandwidth of approximately 200 meV and Hubbard $U$ of 1.2 eV \cite{PhysRevX.13.041049}. The localized electrons within the half-filled flat band are predominantly located in the center of the Nb$_3$ clusters \cite{grytsiuk2024nb3cl8}. Monolayer graphene (MLG) introduces dispersive bands into the system. Density functional theory (DFT) calculations (see methods  \cite{supp} and  \cite{QE-2009,QE-2017}) reveal a near-ideal band alignment (Fig. \ref{F1}b), with graphene's slight \textit{p}-doping preserving the flat band's half-filling. Crucially, interfacial hybridization leads to the opening of a hybridization gap on the order of several tens of meV (inset of Fig. \ref{F1}b), a hallmark of coherent Kondo screening despite DFT's inherent limitations in modeling strong correlations. This vdW heterostructure effectively embodies the essential ingredients of the periodic Anderson model: localized moments coupled with metallic states through spatially resolved orbital overlap.

In investigating the  {phase space of the coupled bilayer}, we fabricated dual-gated heterostructure devices (Fig. \ref{F1}c and Fig. \ref{F1}d), without intentional twist control, given the inherent lattice mismatch between Nb$_3$Cl$_8$ (6.833 \AA) and MLG (2.468 \AA) \cite{Liang_Chen_Wang_Jia_Lu_Xie_Cai_Wang_Meng_Liu_2022}. The observed phenomena exhibit consistency across all five devices, with another representative case illustrated in Fig. S4 \cite{supp}.  {The dual-gate geometry allows the carrier density and the out-of-plane electric field to be tuned independently. In practice, top-gate sweeps at 2 K show negligible hysteresis, whereas back-gate sweeps mediated through Nb$_3$Cl$_8$ exhibit an intriguing and unconventional hysteresis (Fig. S5 \cite{supp}). This unconventional hysteresis is likely related to correlated charge-reconfiguration processes and will be investigated in a separate study. In the present work, we therefore focus on the top-gate-tunable high-resistance regime at fixed $V_{BG}=0$ V, where the transport signatures associated with interfacial hybridization can be analyzed reproducibly.} 

 {Figure \ref{F1}e reveals gate-tunable correlation effects through the temperature-dependent resistance measured at $V_{\rm{TG}}=-1$ V and $V_{\rm{TG}}=4$ V. Below 10 K, both curves are well described by the Fermi-liquid form $R=R_0+AT^2$  \cite{kadowaki1986universal}, indicating that electron–electron scattering provides the dominant temperature-dependent contribution in this low-temperature regime, while phonon scattering is strongly suppressed  \cite{hwang2008acoustic}. Importantly, the value of $A^{1/2}$ at $V_{\rm{TG}}=4$ V is nearly one order of magnitude larger than that at $V_{\rm{TG}}=-1$ V, indicating an enhanced quasiparticle effective mass in the positive-gate regime. Paradoxically, the 2 K resistance value at $V_{\rm{TG}}=4$ V is approximately twice that at $V_{\rm{TG}}=-1$ V. In a conventional monolayer-graphene picture, this would imply a lower carrier density and a smaller graphene-like effective mass; however, the much larger $T^2$ coefficient at $V_{\rm{TG}}=4$ V reveals an enhanced quasiparticle effective mass, demonstrating a clear departure from pristine graphene behavior, and supports a gate-induced reconstruction of the interfacial electronic states.}

\textit{Band structure and the emergent flat band-}-To unravel this discrepancy and delve into interfacial band reconstruction, we conducted magneto-transport measurements that unveiled gate-tunable quantum oscillations. Figure \ref{F2}a charts the symmetrized longitudinal resistance ($R_{\rm{xx}}$) across $V_{\rm{TG}}$ and perpendicular magnetic field, delineating two distinct regimes demarcated by red dashed lines. Within the domain of the small $V_{\rm{TG}}$ ($V_{\rm{TG}}<$0 V, regime II), the Landau fan diagram’s construction reveals linear profiles akin to conventional MLG behavior. Conversely, in the large top gate region (regime I), the Landau levels exhibit a steeper trajectory with discernible minor kinks. The pronounced transition in the Landau levels’ slope as well as in the low-field Hall signal (Fig. S6 \cite{supp}) across the boundary signifies a distinct charge transfer into states beyond graphene’s Dirac cone in regime I  \cite{wang2022quantum, yang2023unconventional}. The decreasing of SdH oscillations periods with increasing gate voltage indicates that the graphene layer remains hole-doped over this gate range, in contrast to pristine MLG devices in which the charge-neutrality point lies close to zero top-gate voltage (Fig. S7 \cite{supp}). The field-tilted phase boundary (red dashed lines in Fig. \ref{F2}a) underscores the field-responsive nature of band alignment. Crucially, the hybrid states’ high density of states (DOS) enforces persistent \textit{p}-doping, effectively locking the system away from charge neutrality — a stark departure from MLG’s gate-response paradigm. 

This charge transfer phenomenon is indicative of the evolution of the DFT band structure, wherein the Fermi level ($E_{\rm{F}}$) initially rises from below the flat band (pure graphene-dominated transport) and eventually reaches and stabilizes at the flat band. However, the model predicts a complete depletion of Nb$_3$Cl$_8$'s half-filled flat band in regime II, resulting in hole-doping with a carrier density ($\sim10^{14}$ cm$^{-2}$) that far exceeds both top gate-induced and Shubnikov-de Haas (SdH) oscillation-derived densities ($\sim10^{12}$ cm$^{-2}$, see Fig. S8 \cite{supp}). This paradox can be resolved through the concept of orbital-selective Mott correlations \cite{chen2024emergent}, which involves incorporating Nb$_3$Cl$_8$'s large and screening-insensitive Hubbard $U$ ($\sim1.2$ eV) \cite{grytsiuk2024nb3cl8}, along with interlayer hybridization. Initially, isolated components exhibit MLG's Dirac cone and Nb$_3$Cl$_8$'s flat band (Fig. \ref{F2}b, upper panel). Introducing interlayer hybridization alone produces the DFT calculated band structure (Fig. \ref{F2}b, middle left panel), while only activating on-site Hubbard $U$ produces split Hubbard bands coexisting with MLG's Dirac cone (Fig. \ref{F2}b, middle right panel). It is imperative to acknowledge that the concurrent hybridization and Hubbard $U$ redistribute quasiparticle spectral weight, coalescing the original two Hubbard bands into three reconstructed bands of lower incoherent, middle hybrid, and upper incoherent bands (Fig. \ref{F2}b, bottom) near the K point. The coherence of the hybridized middle band (with the itinerant bands) drives the interfacial many-body reconstruction, manifesting as either a   {coherent hybridized metal} \cite{de2008t, pepin2007kondo} or a Kondo insulator \cite{schafer2019quantum} contingent on $E_{\rm{F}}$'s position relative to the hybridization gap. 

This low-energy band configuration (Fig. \ref{F2}b, pink dashed box) aligns with DFT outcomes but exhibits finite quasiparticle weight in the middle coherent band at $E_{\rm{F}}$. Notably, the lower band retains graphene-like dispersion due to MLG's high Fermi velocity, which is consistent with conventional MLG transport behavior observed in regime II. Increasing $V_{\rm{TG}}$ drives the Fermi level from the lower band to the middle band, marking coherent charge transfer into hybridized states. The band-resolved analysis quantitatively explains both the observed interfacial charge transfer and MLG's low carrier density, establishing gate-controlled hybridization as the mechanism underpinning correlated electron physics in the heterostructure.

\begin{figure}
\centering
\includegraphics[width=1.0\linewidth]{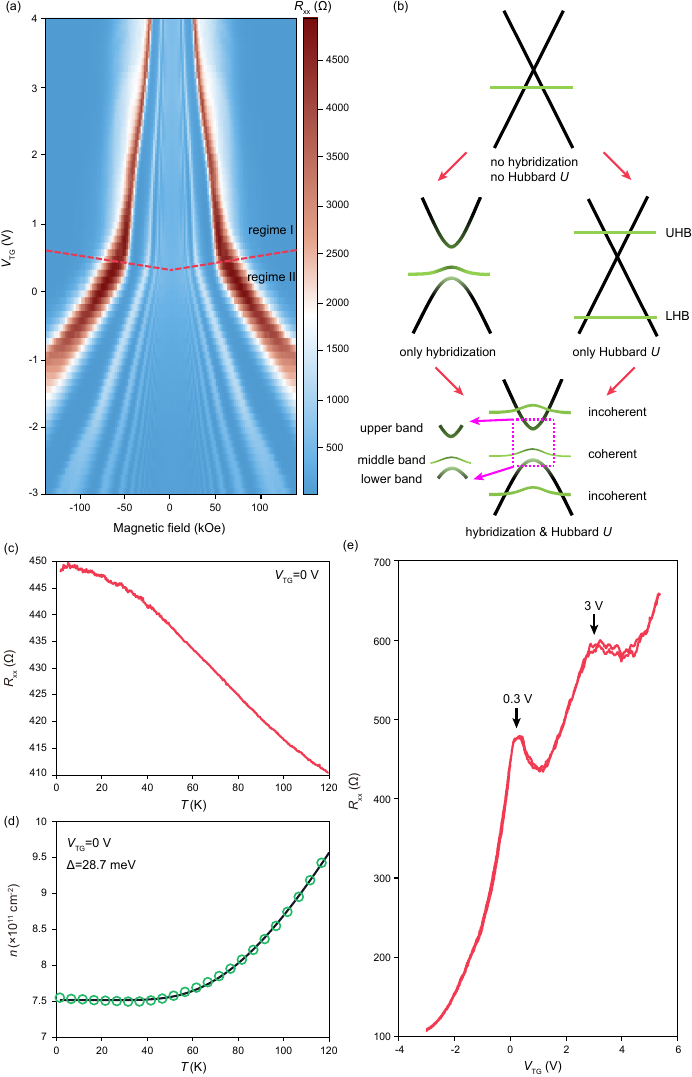}
\caption{\textbf{Electronic band structure and transport properties of the Nb$_3$Cl$_8$/MLG heterostructure.} \textbf{a,} Resistance \textit{versus} $V_{\rm{TG}}$ and magnetic field at 2 K. The red dashed lines demarcate two regimes. 
\textbf{b,} Schematic band structure evolution with interlayer hybridization and Hubbard $U$. The MLG orbit and the Nb$_3$Cl$_8$ orbit are denoted as black and green, respectively.
\textbf{c,} Temperature-dependent resistance at $V_{\rm{TG}}=0$ V. Insulating behavior confirms a hybridization gap.
\textbf{d,} Temperature-dependent carrier density (green circles) at  $V_{\rm{TG}}=0$ V extracted from temperature-dependent Hall measurements. The Fermi-Dirac distribution fit (black line) extracts a 28.7 meV band gap.
\textbf{e,} $V_{\rm{TG}}$-dependent resistance at 2 K and without a magnetic field. Peaks at $V_{\rm{TG}}=0.3$ V and $V_{\rm{TG}}=3$ V (arrows) mark phase boundaries.
}
\label{F2}
\end{figure}

The band reconstruction scenario predicts a hybridization gap near the phase boundary, a key signature of interlayer hybridization coherence. Experimentally, an anomalous insulator-like $R$-$T$ curve emerges at $V_{\rm{TG}}=0$ V (Fig. \ref{F2}c), exhibiting negative $dR/dT$ above 5 K, distinct from metallic responses at other gate voltages either above or below the phase boundary (Fig. \ref{F1}e). The insulator-like behavior indicates a gap at the phase boundary. Temperature-dependent Hall measurements reveal thermally activated carriers across this gap (Fig. \ref{F2}d), 
with density evolution well-described by $n=n_0+n_1/(1+$exp$(\Delta/kT))$, where $n_0$ and $n_1$ represent residual carrier density and the effective DOS at the band edge, respectively. The fitted gap size $\Delta\approx 28.7$ meV confirms strong graphene-flat band hybridization. Two sharp resistance peaks in the $R$-$V_{\rm{TG}}$ profile at 2 K under zero magnetic field (Fig. \ref {F2}e) demarcate insulating regimes: one near charge transfer ($V_{\rm{TG}}=0.3$ V) and another signaling middle-band complexity ($V_{\rm{TG}}=3$ V). The discrepancy in $V_{\rm{TG}}$ between the insulating $R$-$T$ curve ($V_{\rm{TG}}=0$ V) and the peak resistance $R$-$V_{\rm{TG}}$ curve ($V_{\rm{TG}}=0.3$ V) may be attributable to residual carriers. The extracted residual carrier density is substantially large ($n_0 \approx 7.5\times10^{11}$ cm$^{-2}$), dominating low-temperature transport at the charge transfer boundary via in-gap states, causing marked instability in the $R$-$T$ curve at 0 V top gate below 50 K (Fig. S9 \cite{supp}), which is influenced by historical factors and electric noise on the top gate. Notably, the manifestation of in-gap states echoes phenomena observed in Kondo insulators \cite{riseborough2003collapse, pietrus2008kondo}.  {Despite the high quality of our device, as evidenced by the clear SdH oscillations, the finite residual carrier density suggests that the hybridization gap is not spatially uniform across the entire heterostructure. We attribute this effect to intrinsic modulation of the interlayer hybridization caused by the lattice incommensurability between Nb$_3$Cl$_8$ and MLG. Different local stacking configurations produce variations in the bare interlayer coupling, and the graphene wave-function structure can further suppress the coupling in selected local registries. As a result, strongly hybridized regions, where the local gap pins the Fermi level and contributes activated behavior, coexist microscopically with weakly hybridized regions that retain residual graphene-like carriers. Because this modulation is intrinsic to the incommensurate interface and occurs on a length scale much smaller than the device size, macroscopic transport measures a reproducible ensemble-averaged response rather than uncontrolled device disorder. This picture reconciles the large residual carrier density with the thermally activated Hall response and the reproducibility observed across multiple devices.}

 {The coexistence of strongly and weakly coupled regions is further confirmed by temperature-dependent SdH measurement within the two phases. Figure S10 displays the temperature-dependent SdH oscillation at $V_{\rm{TG}}=4$ V and $V_{\rm{TG}}=-1$ V after removing a cubic polynomial background. The extracted carrier cyclotron masses are 0.0088 $m_e$ at $V_{\rm{TG}}=4$ V and 0.0165 $m_e$ at $V_{\rm{TG}}=-1$ V. This result shows consistency with the theoretical prediction of bare MLG Dirac fermions: $m^*=\hbar k_F/v_F=\hbar\sqrt{\pi n}/v_F$. According to the corresponding Hall density (Fig. S6 \cite{supp}), the predicted effective mass is 0.010 $m_e$ and 0.023 $m_e$ for $V_{\rm{TG}}=4$ V and $V_{\rm{TG}}=-1$ V, respectively, deviating from experimental value by less than 30\%, which may result from Fermi velocity renormalization due to electron-electron interactions. These carrier cyclotron masses do not show a large enhancement, contradicting the clear effective mass enhancement that derived from the $R-T$ relation. This discrepancy could be understood from the channel-selective detection of the two methods: while the temperature-dependent resistance samples both regions equally, the SdH oscillation responds predominantly to channel with high mobility and low effective mass. According to the Lifshitz–Kosevich formula, the amplitude of SdH oscillation is proportional to $\exp(-\beta T_D)\times\beta T/\sinh(\beta T)$, where $\beta=2 \pi^2 k_B m^*/\hbar e B$ and $T_D=\hbar/2\pi k_B\tau_q$; $\tau_q$ is the quantum scattering time. This formula indicates that the amplitude is strongly suppressed by the large carrier effective mass and low scattering time. Such channel-selectivity of SdH oscillation is well established in multi-channel heavy fermion compounds. For example, in URu$_2$Si$_2$, the angle-dependent SdH measurement misses an electron Fermi surface with effective mass more than 30 $m_e$ \cite{URu2Si2}. In our heterostructure, the spatial modulation of interlayer hybridization divides the entire heterostructure into weakly and strongly coupled regions. The electrons gather high effective mass and additional scattering from Nb$_3$Cl$_8$ orbits in the strongly coupled region, while retain their intrinsic low effective mass and high mobility in the weakly coupled region. Consequently, the temperature-dependent SdH oscillation is dominated by the weakly coupled region, in which the electron effective mass resembles the intrinsic value. Therefore, the temperature dependent SdH only reflects the weakly coupled region, while the temperature-dependent resistance reveals information of strongly coupled region.}

 {Taken together, the gate-dependent charge transfer, activated Hall response, hybridization-gap scale, and mass enhancement provide mutually consistent evidence for interfacial correlated band reconstruction and support the assignment of a    {interlayer coherence} in the Nb$_3$Cl$_8$/MLG heterostructure.} This places the system in a mixed-valence regime \cite{varma1976mixed}, where the filling of Nb$_3$Cl$_8$’s localized flat band is continuously tuned by the top gate. The observed gate-tunable effective electron mass dichotomy stems from contrasting flat band contributions: The middle band’s dominant flat band character under positive $V_{\rm{TG}}$ drives electron effective mass renormalization, while the lower band’s graphene-like dispersion at negative $V_{\rm{TG}}$ yields conventional behavior. This spectral weight asymmetry directly links band dispersion to emergent quantum phases.

\begin{figure}[ht!]
\centering 
\includegraphics[width=0.5\textwidth]{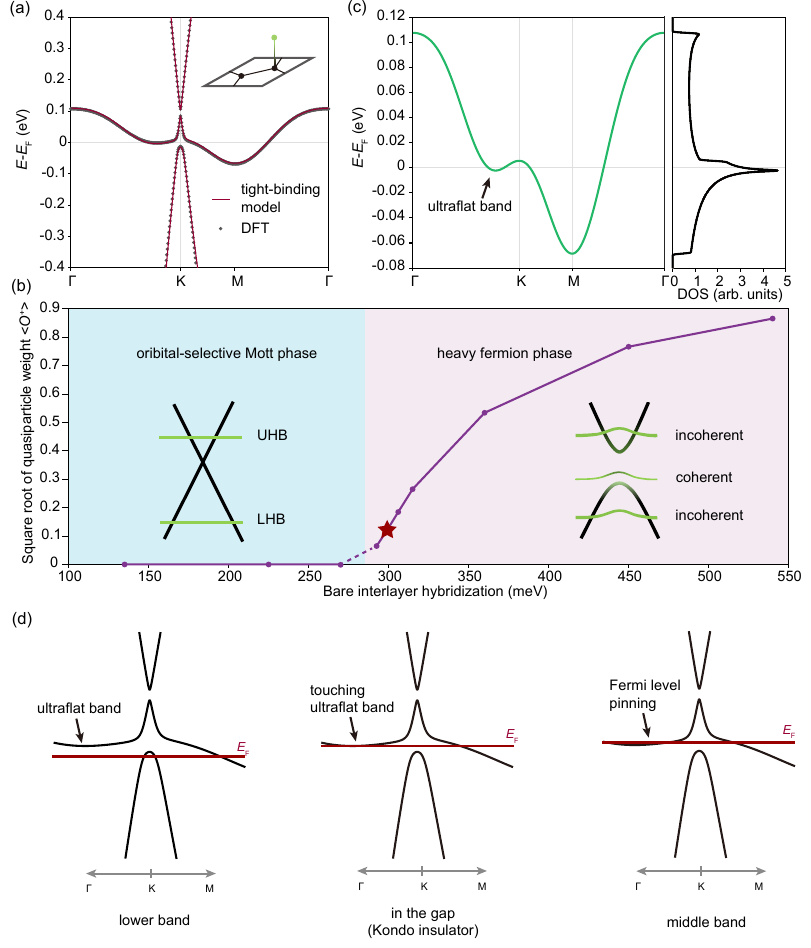}
\caption{\label{3}\textbf{Microscopic mechanism of correlated phenomena in the Nb$_3$Cl$_8$/MLG heterostructure.}
 \textbf{a,} DFT-calculated (gray dots) and tight-binding model calculated (red lines) of the low-energy band structure of the heterostructure. Inset: Schematic of the tight-binding model geometry (black dots: MLG carbon atoms; green dot: Nb$_3$ cluster).
 \textbf{b,} Phase diagram from SSMF calculations. The order parameter $\langle O^+\rangle$ transitions abruptly from zero to a finite value across the phase transition. The star denotes the system's experimental position, determined by matching the observed hybridization gap.
 \textbf{c,} Tight-binding band structure (left panel) and DOS (right panel) for isolated Nb$_3$Cl$_8$.
 \textbf{d,} Band structure and Fermi level evolution near the charge transfer phase boundary. The heterostructure transitions from regime II (left) to regime I (right) as the Fermi level rises across the hybridization gap.
 }
\label{F3}
\end{figure}

\textit{Slave spin mean-field calculation and microscopic mechanism-}-To elucidate the microscopic picture of the  {experimental observation}, including the emergent coherent flat band, charge transfer behavior, and distinct electronic regimes across the charge transfer phase boundary, we employ a slave spin mean-field (SSMF) calculation tailored to the Nb$_3$Cl$_8$/MLG heterostructure, a method that captures key mechanisms of   {PAM} (Fig. \ref{F3}a and see details in Supplementary Note II \cite{supp}).
 While standard DFT fails to describe correlation-driven phenomena, the SSMF framework renormalizes hybridization, bandwidth, and onsite energies, enabling direct comparison with experimental trends. 

As the bare hybridization strength increases, the SSMF calculation unveils a phase transition from an orbital-selective Mott phase (OSMP) to a  {heavy} fermion phase, characterized by the emergence of a quasiparticle weight (Fig. \ref{F3}b). In the   {phase space with finite quasiparticle weight}, the theory predicts a coherent flat band that hybridizes with the dispersive bands of MLG, opening a gap at the K point, in alignment with our experimental findings (the position determined by the gap size is marked by the red star in the phase diagram). The system's {finite quasiparticle weight} is further corroborated by the enhanced electron effective mass of the middle band (Supplementary Note III and Fig. S2 \cite{supp}, and reference  \cite{guan2017band,ariel2013electron} therein), aligning with transport data. Our picture is also compatible with recent STM/STS measurements on MLG/Nb$_3$Cl$_8$ heterostructures, in which Hubbard-band features were clearly resolved  \cite{liu2025direct}. The absence of a pronounced middle coherent band in the reported graphene-side tunneling spectra does not necessarily contradict our interpretation. In the present scenario, the system lies close to the orbital-selective-Mott/heavy-fermion boundary, where only limited spectral weight is transferred to the middle coherent band. Moreover, because the STM tip probes the exposed graphene surface, the measured tunneling conductance is expected to be dominated by the graphene-projected density of states. The relatively weak hybridization feature may therefore be partially masked by the intrinsic phonon-assisted tunneling dip of monolayer graphene. This layer-selective tunneling geometry naturally explains why transport can be sensitive to interfacial hybridization even when the middle band is not sharply resolved in graphene-side STS.


\begin{figure}[ht!]
\centering 
\includegraphics[width=0.5\textwidth]{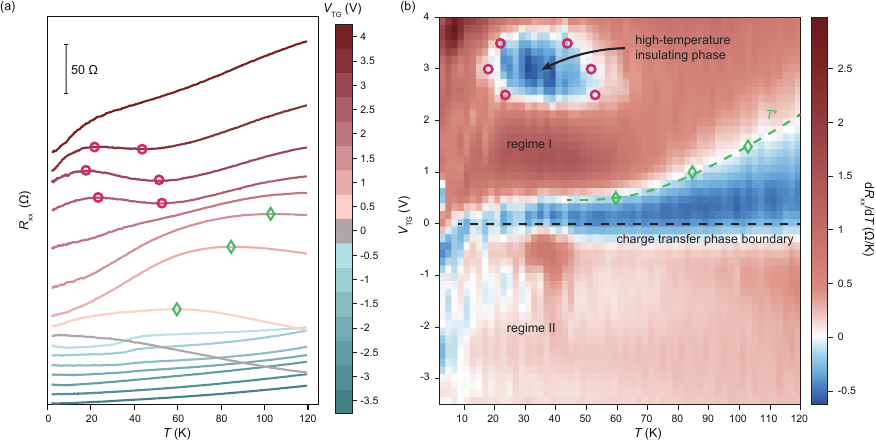}
\caption{\label{4}\textbf{Gate-temperature-tunable phase diagram of the Nb$_3$Cl$_8$/MLG heterostructure.}
 \textbf{a,} The shifted temperature-dependent resistance at various $V_{\rm{TG}}$ from $-$3.5 V to 4 V.  
 \textbf{b,} $dR/dT$ map across $V_{\rm{TG}}$ and $T$. The curves are smoothed as detailed in the method section with $\Delta T$ = 2 K and interpolated along $V_{\rm{TG}}$ at an interval of 0.15 V. Regime I and regime II align with Fig. \ref{F2}a.}
\label{F4}
\end{figure}

Although SSMF successfully reproduces most experimental observations, it is limited by the underestimation of bandwidth due to neglected quantum fluctuations. Despite that, the theory offers a qualitatively sound understanding of the microscopic picture by manually addressing these discrepancies. Initially, according to the mean-field constraint condition, the $E_{\rm{F}}$ consistently intersects the emergent coherent band. However, the Nb$_3$Cl$_8$'s substantial Hubbard $U$ triggers robust onsite energy renormalization, which ultimately leads to the coherent band tracking the $E_{\rm{F}}$ instead of pinning it. This facilitates efficient gate-tuning of MLG's carrier density, a pivotal feature absent in DFT. Furthermore, a pronounced DOS peak near the half-filling of the Nb$_3$Cl$_8$ band (Fig. \ref{3}c) governs charge transfer, akin to non-interacting systems. All these facets elucidate the scenario near the charge transfer boundary (Fig. \ref{F3}d). Originating from the lower band phase, with the increase of $V_{\rm{TG}}$, the $E_{\rm{F}}$ ascends, consequently elevating the flat band due to onsite energy renormalization, enabling carriers to populate the MLG bands. As $V_{\rm{TG}}\approx0$ V, the system transitions into the Kondo insulator phase, where the $E_{\rm{F}}$ resides within the gap, engendering insulating behavior in the heterostructure. The $E_{\rm{F}}$ also touches the ultra-flat band, contributing to the DOS peak near zero top gate, ushering the system into regime I. Subsequently, as $E_{\rm{F}}$ continues its ascent, the system traverses into another metallic phase, accompanied by an enhancement in electron effective mass. Significantly, the three-band representation near the K point, together with the fact that MLG governs transport by its superior mobility, rationalizes the observed metal-insulator transition and anomalous temperature-dependent behavior. By bridging correlations, band properties, and gate response, the SSMF framework {identifies Nb$_3$Cl$_8$/MLG at the crossover between the OSMP and the heavy fermion phase, with non-zero quasiparticle weight} embodying intertwined quantum phases.

\textit{$V_{\rm{TG}}$ \textit{versus} $T$ phase diagram-}-Our comprehensive phase mapping uncovers rich correlation-driven phenomena in the heterostructure. Figure \ref{F4}a displays the temperature-dependent resistance measured at varied $V_{\rm{TG}}$. The low-temperature data align with our theoretical model, exhibiting two distinct regimes: (i) At negative $V_{\rm{TG}}$, the $E_{\rm{F}}$ intersects the lower band's graphene-like dispersion, yielding in marginal resistance variation from 2-20 K — a hallmark of weak correlations. (ii) Positive $V_{\rm{TG}}$ shifts $E_{\rm{F}}$ into the hybridized middle band, driving pronounced resistance increases with temperature (\textit{e.g.}, $+$10.6 $\Omega$ at $V_{\rm{TG}}$=2 V, 2-20 K). Near the charge transfer boundary ($V_{\rm{TG}}\approx$0 V), anomalous insulating behavior emerges (negative $dR/dT$ above 5 K), signaling a Kondo insulator scenario in our microscopic model. The derived $dR/dT$ map across the $V_{\rm{TG}}$-$T$ parameter space (Fig. \ref{F4}b) reveals critical phase boundaries underpinning the system's quantum complexity. The charge transfer boundary demarcates a prominent insulating zone, corresponding to the hybridization gap discussed earlier, which dominates transport. At elevated temperatures, the insulating region extends into regime I at higher $V_{\rm{TG}}$ (green dashed line, Fig. \ref{F4}b), a signature  {similar to that }of  {heavy} fermion metals characterized by low carrier densities \cite{das2018magnetic, wang2017heavy}. Centered at 40 K and $V_{\rm{TG}}=3$ V, a high-temperature insulator-like phase emerges, coinciding with the resistance peak in Fig. \ref{F2}e, and may originate from another kind of incoherent Kondo scattering. In regime II at $V_{\rm{TG}}$ from $-2$ V to 0 V, we observe a rapid resistance increase near 40 K. At strongly negative $V_{\rm{TG}}$ ($-3$ V to $-2$ V), a low-temperature ($<10$ K) insulating pocket forms. These phases collectively suggest competing electronic orders. 

Strikingly, in-plane magnetic fields induce a metal-insulator transition in the middle band ($V_{\rm{TG}}=0.5$ V) while leaving the lower band ($V_{\rm{TG}}=-1$ V) unaffected (Fig. S11 \cite{supp}), which is direct evidence of band-selective quantum phases. While mean-field theories capture Kondo coherence and the hybridization gap, the multiphase complexity, including competing insulating regimes and new incoherent Kondo scattering, demands advanced frameworks incorporating magnetic frustration and lattice-mismatch effects. This system’s gate-tunable phase multiplicity establishes it as a unique platform for probing quantum criticality near competing orders.

\textit{Conclusion-}- {The Nb$_3$Cl$_8$/MLG heterostructure provides an electrically tunable platform for studying  {localized-itinerant electron coupling} in an incommensurate vdW interface. The observed hybridization gap of approximately 30 meV, gate-dependent charge transfer, Fermi-liquid coefficient enhancement, and band-selective transport behavior are consistently captured by a slave-spin mean-field framework based on localized Nb$_3$Cl$_8$ states coupled to itinerant graphene carriers. Although residual light-carrier channels and spatially nonuniform hybridization moderate the apparent mass enhancement, they are naturally expected for an incommensurate interface and are incorporated into the microscopic interpretation.  Beyond this specific material combination, our results suggest that coupling a Mott-insulating layer to a tunable itinerant-electron system can provide a useful strategy for exploring correlated interfacial states in vdW heterostructures. By relaxing strict lattice-matching constraints, this approach may be extended to other itinerant layers, including spin-orbit-coupled transition-metal dichalcogenides  \cite{yang2017strong} and graphitic systems with nontrivial band topology  \cite{han2024correlated}. Such platforms could offer opportunities to investigate low-dimensional quantum criticality, competing correlated phases, and electrically tunable quasiparticle reconstruction in engineered quantum materials.}

\begin{acknowledgments}
\textit{Acknowledgments-}-This work was supported by the National Natural Science Foundation of China (No. 12425402) and the National Key R\&D Program of China (No. 2025YFA1411002 and No. 2022YFA1203902). K.W. and T.T. acknowledge support from the JSPSKAKENHI (No. 21H05233 and No. 23H02052) and World Premier International Research Center Initiative (WPI), MEXT, Japan. 

\end{acknowledgments}
\textit{Authors contributions-}-Y.Y. and Y.G. conceived the project. Y.G. fabricated the devices, performed the transport measurements, analyzed the data, and conducted the theoretical calculation with the help of W.Z., F. Y., X.H., and H.X.. Z.M. synthesized the Nb$_3$Cl$_8$ crystals under the supervision of Y.S.. K.W. and T.T. grew the \textit{h}-BN bulk crystals. Y.G. and Y.Y. drafted the manuscript. All authors discussed the results and contributed to the manuscript.
\bibliographystyle{apsrev4-2-titles}
\bibliography{Reference} 

@article{dzero2016topological,
  title={Topological {Kondo} insulators},
  author={Dzero, Maxim and Xia, Jing and Galitski, Victor and Coleman, Piers},
  journal={Annual Review of Condensed Matter Physics},
  volume={7},
  number={1},
  pages={249--280},
  year={2016},
  publisher={Annual Reviews}
}

@article{jiao2020chiral,
  title={Chiral superconductivity in heavy-fermion metal {UTe}$_2$},
  author={Jiao, Lin and Howard, Sean and Ran, Sheng and Wang, Zhenyu and Rodriguez, Jorge Olivares and Sigrist, Manfred and Wang, Ziqiang and Butch, Nicholas P and Madhavan, Vidya},
  journal={Nature},
  volume={579},
  number={7800},
  pages={523--527},
  year={2020},
  publisher={Nature Publishing Group UK London}
}

@article{stewart2001non,
  title = {Non-Fermi-liquid behavior in $d$- and $f$-electron metals},
  author = {Stewart, G. R.},
  journal = {Reviews of Modern Physics},
  volume = {73},
  issue = {4},
  pages = {797--855},
  numpages = {0},
  year = {2001},
  month = {Oct},
  publisher = {American Physical Society},
  doi = {10.1103/RevModPhys.73.797},
  url = {https://link.aps.org/doi/10.1103/RevModPhys.73.797}
}

@article{vavno2021artificial,
  title={Artificial heavy fermions in a van der {Waals} heterostructure},
  author={Va{\v{n}}o, Viliam and Amini, Mohammad and Ganguli, Somesh C and Chen, Guangze and Lado, Jose L and Kezilebieke, Shawulienu and Liljeroth, Peter},
  journal={Nature},
  volume={599},
  number={7886},
  pages={582--586},
  year={2021},
  publisher={Nature Publishing Group UK London}
}

@article{zhao2023gate,
  title={Gate-tunable heavy fermions in a moir{\'e} {Kondo} lattice},
  author={Zhao, Wenjin and Shen, Bowen and Tao, Zui and Han, Zhongdong and Kang, Kaifei and Watanabe, Kenji and Taniguchi, Takashi and Mak, Kin Fai and Shan, Jie},
  journal={Nature},
  volume={616},
  number={7955},
  pages={61--65},
  year={2023},
  publisher={Nature Publishing Group UK London}
}

@article{PhysRevX.13.041049,
  title = {Discovery of a Single-Band {Mott} {Insulator} in a van der {Waals} {Flat-Band} {Compound}},
  author = {Gao, Shunye and Zhang, Shuai and Wang, Cuixiang and Yan, Shaohua and Han, Xin and Ji, Xuecong and Tao, Wei and Liu, Jingtong and Wang, Tiantian and Yuan, Shuaikang and Qu, Gexing and Chen, Ziyan and Zhang, Yongzhao and Huang, Jierui and Pan, Mojun and Peng, Shiyu and Hu, Yong and Li, Hang and Huang, Yaobo and Zhou, Hui and Meng, Sheng and Yang, Liu and Wang, Zhiwei and Yao, Yugui and Chen, Zhiguo and Shi, Ming and Ding, Hong and Yang, Huaixin and Jiang, Kun and Li, Yunliang and Lei, Hechang and Shi, Youguo and Weng, Hongming and Qian, Tian},
  journal = {Phys. Rev. X},
  volume = {13},
  issue = {4},
  pages = {041049},
  numpages = {9},
  year = {2023},
  month = {Dec},
  publisher = {American Physical Society},
  doi = {10.1103/PhysRevX.13.041049},
  url = {https://link.aps.org/doi/10.1103/PhysRevX.13.041049}
}

@article{chen2024emergent,
  title={Emergent flat band and topological {Kondo} semimetal driven by orbital-selective correlations},
  author={Chen, Lei and Xie, Fang and Sur, Shouvik and Hu, Haoyu and Paschen, Silke and Cano, Jennifer and Si, Qimiao},
  journal={Nature Communications},
  volume={15},
  number={1},
  pages={5242},
  year={2024},
  publisher={Nature Publishing Group UK London}
}

@article{grytsiuk2024nb3cl8,
  title={{Nb}$_3${Cl}$_8$: a prototypical layered {Mott-Hubbard} insulator},
  author={Grytsiuk, Sergii and Katsnelson, Mikhail I and Loon, Erik GCP van and R{\"o}sner, Malte},
  journal={npj Quantum Materials},
  volume={9},
  number={1},
  pages={8},
  year={2024},
  publisher={Nature Publishing Group UK London}
}

@article{kadowaki1986universal,
  title={Universal relationship of the resistivity and specific heat in heavy-fermion compounds},
  author={Kadowaki, K and Woods, SB},
  journal={Solid state communications},
  volume={58},
  number={8},
  pages={507--509},
  year={1986},
  publisher={Elsevier}
}

@article{wang2022quantum,
  title={Quantum {Hall} phase in graphene engineered by interfacial charge coupling},
  author={Wang, Yaning and Gao, Xiang and Yang, Kaining and Gu, Pingfan and Lu, Xin and Zhang, Shihao and Gao, Yuchen and Ren, Naijie and Dong, Baojuan and Jiang, Yuhang and others},
  journal={Nature Nanotechnology},
  volume={17},
  number={12},
  pages={1272--1279},
  year={2022},
  publisher={Nature Publishing Group UK London}
}

@article{riseborough2003collapse,
  title={Collapse of the coherence gap in {Kondo} semiconductors},
  author={Riseborough, Peter S},
  journal={Physical Review B},
  volume={68},
  number={23},
  pages={235213},
  year={2003},
  publisher={APS}
}

@article{pietrus2008kondo,
  title={Kondo-hole conduction in the {La}-doped {Kondo} insulator {Ce}$_3$ {Bi}$_4$ {Pt}$_3$},
  author={Pietrus, T and v. L{\"o}hneysen, H and Schlottmann, P},
  journal={Physical Review B},
  volume={77},
  number={11},
  pages={115134},
  year={2008},
  publisher={APS}
}

@article{Liang_Chen_Wang_Jia_Lu_Xie_Cai_Wang_Meng_Liu_2022,
  title={A universal model for accurately predicting the formation energy of inorganic compounds},
  author={Liang, Yingzong and Chen, Mingwei and Wang, Yanan and Jia, Huaxian and Lu, Tenglong and Xie, Fankai and Cai, Guanghui and Wang, Zongguo and Meng, Sheng and Liu, Miao},
  DOI={10.1007/s40843-022-2134-3},
  journal={Science China Materials},
  volume={66},
  number={1},
  pages={343--351},
  year={2023},
  publisher={Springer}
}

@article{yang2023unconventional,
  title={Unconventional correlated insulator in {CrOCl}-interfaced {Bernal} bilayer graphene},
  author={Yang, Kaining and Gao, Xiang and Wang, Yaning and Zhang, Tongyao and Gao, Yuchen and Lu, Xin and Zhang, Shihao and Liu, Jianpeng and Gu, Pingfan and Luo, Zhaoping and others},
  journal={Nature Communications},
  volume={14},
  number={1},
  pages={2136},
  year={2023},
  publisher={Nature Publishing Group UK London}
}

@article{varma1976mixed,
  title = {Mixed-valence compounds},
  author = {Varma, C. M.},
  journal = {Reviews of Modern physics},
  volume = {48},
  issue = {2},
  pages = {219--238},
  numpages = {0},
  year = {1976},
  month = {Apr},
  publisher = {American Physical Society},
  doi = {10.1103/RevModPhys.48.219},
  url = {https://link.aps.org/doi/10.1103/RevModPhys.48.219}
}

@article{yang2017strong,
  title={Strong electron-hole symmetric {Rashba} spin-orbit coupling in graphene/monolayer transition metal dichalcogenide heterostructures},
  author={Yang, Bowen and Lohmann, Mark and Barroso, David and Liao, Ingrid and Lin, Zhisheng and Liu, Yawen and Bartels, Ludwig and Watanabe, Kenji and Taniguchi, Takashi and Shi, Jing},
  journal={Physical Review B},
  volume={96},
  number={4},
  pages={041409},
  year={2017},
  publisher={APS}
}

@article{han2024correlated,
  title={Correlated insulator and {Chern} insulators in pentalayer rhombohedral-stacked graphene},
  author={Han, Tonghang and Lu, Zhengguang and Scuri, Giovanni and Sung, Jiho and Wang, Jue and Han, Tianyi and Watanabe, Kenji and Taniguchi, Takashi and Park, Hongkun and Ju, Long},
  journal={Nature Nanotechnology},
  volume={19},
  number={2},
  pages={181--187},
  year={2024},
  publisher={Nature Publishing Group UK London}
}

@article{QE-2017,
  author={P Giannozzi and O Andreussi and T Brumme and O Bunau and M Buongiorno Nardelli
  and M Calandra and R Car and C Cavazzoni and D Ceresoli and M Cococcioni and N Colonna
  and I Carnimeo and A Dal Corso and S de Gironcoli and P Delugas and R A DiStasio Jr and A Ferretti
  and A Floris and G Fratesi and G Fugallo and R Gebauer and U Gerstmann and F Giustino and T Gorni
  and J Jia and M Kawamura and H-Y Ko and A Kokalj and E Küçükbenli and M Lazzeri and M Marsili
  and N Marzari and F Mauri and N L Nguyen and H-V Nguyen and A Otero-de-la-Roza and L Paulatto
  and S Poncé and D Rocca and R Sabatini and B Santra and M Schlipf and A P Seitsonen
  and A Smogunov and I Timrov and T Thonhauser and P Umari and N Vast and X Wu and S Baroni},
  title={Advanced capabilities for materials modelling with QUANTUM ESPRESSO},
  journal={Journal of Physics: Condensed Matter},
  volume={29},
  number={46},
  pages={465901},
  url={http://stacks.iop.org/0953-8984/29/i=46/a=465901},
  year={2017},
}

@article{QE-2009,
	Author = {Paolo Giannozzi and Stefano Baroni and Nicola Bonini and Matteo Calandra and Roberto Car
 and Carlo Cavazzoni and Davide Ceresoli and Guido L Chiarotti and Matteo Cococcioni and Ismaila Dabo
 and Andrea {Dal Corso} and Stefano de Gironcoli and Stefano Fabris and Guido Fratesi and Ralph Gebauer
 and Uwe Gerstmann and Christos Gougoussis and Anton Kokalj and Michele Lazzeri and Layla Martin-Samos
 and Nicola Marzari and Francesco Mauri and Riccardo Mazzarello and Stefano Paolini and Alfredo Pasquarello
 and Lorenzo Paulatto and Carlo Sbraccia and Sandro Scandolo and Gabriele Sclauzero and Ari P Seitsonen
 and Alexander Smogunov and Paolo Umari and Renata M Wentzcovitch},
	Journal = {Journal of Physics: Condensed Matter},
	Number = {39},
	Pages = {395502},
	Title = {QUANTUM ESPRESSO: a modular and open-source software project for quantum simulations of materials},
	Url = {http://www.quantum-espresso.org},
	Volume = {21},
	Year = {2009}}

@article{das2018magnetic,
  title={Magnetic field driven complex phase diagram of antiferromagnetic heavy-fermion superconductor {Ce}$_3${PtIn}$_{11}$},
  author={Das, Debarchan and Gnida, Daniel and Bochenek, {\L}ukasz and Rudenko, Andriy and Daszkiewicz, Marek and Kaczorowski, Dariusz},
  journal={Scientific Reports},
  volume={8},
  number={1},
  pages={16703},
  year={2018},
  publisher={Nature Publishing Group UK London}
}

@article{wang2017heavy,
  title={Heavy fermion behavior in the quasi-one-dimensional Kondo lattice {CeCo}$_2${Ga}$_8$},
  author={Wang, Le and Fu, Zhaoming and Sun, Jianping and Liu, Min and Yi, Wei and Yi, Changjiang and Luo, Yongkang and Dai, Yaomin and Liu, Guangtong and Matsushita, Yoshitaka and others},
  journal={npj Quantum Materials},
  volume={2},
  number={1},
  pages={36},
  year={2017},
  publisher={Nature Publishing Group UK London}
}

@article{guan2017band,
  title={Band gap opening of graphene by forming a graphene/{PtSe}$_2$ van der {Waals} heterojunction},
  author={Guan, Zhaoyong and Ni, Shuang and Hu, Shuanglin},
  journal={RSC advances},
  volume={7},
  number={72},
  pages={45393--45399},
  year={2017},
  publisher={Royal Society of Chemistry}
}

@INPROCEEDINGS{ariel2013electron,
  author={Ariel, V. and Natan, A.},
  booktitle={2013 International Conference on Electromagnetics in Advanced Applications (ICEAA)}, 
  title={Electron effective mass in graphene}, 
  year={2013},
  volume={},
  number={},
  pages={696-698},
  doi={10.1109/ICEAA.2013.6632334}}

@article{schafer2019quantum,
  title={Quantum criticality in the two-dimensional periodic {Anderson} model},
  author={Sch{\"a}fer, Thomas and Katanin, AA and Kitatani, Motoharu and Toschi, Alessandro and Held, Karsten},
  journal={Physical Review Letters},
  volume={122},
  number={22},
  pages={227201},
  year={2019},
  publisher={APS}
}

@article{de2008t,
  title={T= 0 heavy-fermion quantum critical point as an orbital-selective {Mott} transition},
  author={De Leo, Lorenzo and Civelli, Marcello and Kotliar, Gabriel},
  journal={Physical Review Letters},
  volume={101},
  number={25},
  pages={256404},
  year={2008},
  publisher={APS}
}

@article{pepin2007kondo,
  title={Kondo breakdown as a selective {Mott} transition in the {Anderson} lattice},
  author={P{\'e}pin, C},
  journal={Physical Review Letters},
  volume={98},
  number={20},
  pages={206401},
  year={2007},
  publisher={APS}
}

@article{maksimovic2022evidence,
  title={Evidence for a delocalization quantum phase transition without symmetry breaking in {CeCoIn}$_5$},
  author={Maksimovic, Nikola and Eilbott, Daniel H and Cookmeyer, Tessa and Wan, Fanghui and Rusz, Jan and Nagarajan, Vikram and Haley, Shannon C and Maniv, Eran and Gong, Amanda and Faubel, Stefano and others},
  journal={Science},
  volume={375},
  number={6576},
  pages={76--81},
  year={2022},
  publisher={American Association for the Advancement of Science}
}

@article{paschen2004hall,
  title={Hall-effect evolution across a heavy-fermion quantum critical point},
  author={Paschen, Silke and L{\"u}hmann, Thomas and Wirth, Steffen and Gegenwart, Philipp and Trovarelli, Octavio and Geibel, Christoph and Steglich, Frank and Coleman, Piers and Si, Qimiao},
  journal={Nature},
  volume={432},
  number={7019},
  pages={881--885},
  year={2004},
  publisher={Nature Publishing Group UK London}
}

@article{stockert2011magnetically,
  title={Magnetically driven superconductivity in {CeCu}$_2${Si}$_2$},
  author={Stockert, Oliver and Arndt, J and Faulhaber, Enrico and Geibel, Christoph and Jeevan, Hirale S and Kirchner, S and Loewenhaupt, Michael and Schmalzl, K and Schmidt, W and Si, Qimiao and others},
  journal={Nature Physics},
  volume={7},
  number={2},
  pages={119--124},
  year={2011},
  publisher={Nature Publishing Group UK London}
}

@article{liu2025direct,
  title={Direct evidence of intrinsic {Mott} state and its layer-parity oscillation in a breathing kagome crystal down to monolayer},
  author={Liu, Huanyu and Li, Wenhui and Zhou, Zishu and Qu, Hongbin and Zhang, Jiaqi and Hu, Weixiong and Wen, Chenhaoping and Wang, Ning and Deng, Hao and Li, Gang and others},
  journal={Physical Review Letters},
  volume={135},
  number={7},
  pages={076503},
  year={2025},
  publisher={APS}
}

@article{hwang2008acoustic,
  title={Acoustic phonon scattering limited carrier mobility in two-dimensional extrinsic graphene},
  author={Hwang, EH and Das Sarma, S},
  journal={Physical Review B},
  volume={77},
  number={11},
  pages={115449},
  year={2008},
  publisher={APS}
}

@article{URu2Si2,
author = {Aoki ,Dai and Knebel ,Georg and Sheikin ,Ilya and Hassinger ,Elena and Malone ,Liam and Matsuda, Tatsuma D. and Flouquet ,Jacques},
title = {High-Field {Fermi} Surface Properties in the Low-Carrier Heavy-Fermion Compound {URu}$_2${Si}$_2$},
journal = {Journal of the Physical Society of Japan},
volume = {81},
number = {7},
pages = {074715},
year = {2012},
doi = {10.1143/JPSJ.81.074715},
URL = { 
        https://doi.org/10.1143/JPSJ.81.074715
},
}

@misc{supp,
note={See Supplemental Material at [URL will be inserted by publisher] for methods; low-energy model of Nb$_3$Cl$_8$/MLG heterostructure and calculation of the DOS of Nb$_3$Cl$_8$; slave spin mean-field calculations; estimation of electron effective mass in the middle and lower bands; discussion on potential configurations of band structures; transport phenomena reproducibility; back gate sweeping hysteresis; relation between carrier density and top gate voltage; control experiment of Nb$_3$Cl$_8$/$h$-BN/MLG heterostructure; SdH oscillations at $V_{\rm{TG}}=-0.5$ V; hysteretic $R$-$T$ behavior at $V_{\rm{TG}}=0$ V; temperature-dependent SdH oscillations; in-plane magnetic field-driven metal-insulator-like transition in the middle band.}

}
\end{document}